\documentclass[twocolumn,showpacs,preprintnumbers,amsmath,amssymb,prb]{revtex4}

\usepackage{graphicx}
\usepackage{dcolumn}
\usepackage{bm}
\usepackage{tabularx}
\usepackage{amsmath}

\begin{document}

\title{Nodal gap in iron-based superconductor CsFe$_2$As$_2$ probed by quasiparticle heat transport}

\author{X. C. Hong,$^1$ X. L. Li,$^1$ B. Y. Pan,$^1$ L. P. He,$^1$ A. F. Wang,$^2$ X. G. Luo,$^2$ X. H. Chen,$^2$ and S. Y. Li$^{1,*}$}

\affiliation{$^1$State Key Laboratory of Surface Physics, Department
of Physics, and Laboratory of Advanced Materials, Fudan University,
Shanghai 200433, P. R. China\\
$^2$Hefei National Laboratory for Physical Science at Microscale and
Department of Physics, University of Science and Technology of
China, Hefei, Anhui 230026, China}

\date{\today}

\begin{abstract}
The thermal conductivity of iron-based superconductor CsFe$_2$As$_2$
single crystal ($T_c =$ 1.81 K) was measured down to 50 mK. A
significant residual linear term $\kappa_0/T$ = 1.27 mW K$^{-2}$
cm$^{-1}$ is observed in zero magnetic field, which is about 1/10 of
the normal-state value in upper critical field $H_{c2}$. In low
magnetic field, $\kappa_0/T$ increases rapidly with field. The
normalized $\kappa_0/T(H)$ curve for our CsFe$_2$As$_2$ (with
residual resistivity $\rho_0$ = 1.80 $\mu\Omega$ cm) lies between
the dirty KFe$_2$As$_2$ (with $\rho_0$ = 3.32 $\mu\Omega$ cm) and
the clean KFe$_2$As$_2$ (with $\rho_0$ = 0.21 $\mu\Omega$ cm), which
is consistent with its impurity level. These results strongly
suggest nodal superconducting gap in CsFe$_2$As$_2$, similar to its
sister compound KFe$_2$As$_2$.

\end{abstract}


\maketitle

\section{Introduction}

For the iron-based superconductors,
 \cite{Kamihara,XHChen} one very important issue is the symmetry and structure of their
superconducting gap, \cite{Hirschfeld} which is crucial for
understanding the mechanism of high-temperature superconductivity.
\cite{FaWang} However, after five-year extensive studies, it is
still in a complex situation, mainly due to their multiple
electronic bands. \cite{IIMazin,HDing}

Most of iron-based superconductors have both hole and electron Fermi
surfaces, for example Ba$_{0.6}$K$_{0.4}$Fe$_2$As$_2$. \cite{HDing}
While many of these superconductors show nodeless superconducting
gaps, such as optimally doped BaFe$_2$As$_2$,
\cite{HDing,KTerashima,XGLuo,LDing,Tanatar1} LiFeAs,
\cite{Borisenko,Tanatar2,HKim,KUmezawa} NaFe$_{1-x}$Co$_x$As,
\cite{ZHLiu,SYZhou} and FeTe$_{1-x}$Se$_x$, \cite{JKDong1,KOkazaki1}
some of them manifest nodal superconducting gap, such as
BaFe$_2$(As$_{1-x}$P$_x$)$_2$, \cite{YNakai,KHashimoto1,YZhang1}
Ba(Fe$_{1-x}$Ru$_x$)$_2$As$_2$, \cite{XQiu} LiFeP,
\cite{KHashimoto2} and LaFePO. \cite{Fletcher,Hicks} So far, it is
not conclusive that the nodeless gaps on different Fermi surfacs are
$s_{\pm}$-wave. \cite{Hirschfeld} The nodal gap in those isovalently
P- or Ru-substituted compounds could be accidental nodal $s$-wave on
some Fermi surface, \cite{YZhang1} but its origin is still not very
clear. \cite{KHashimoto2}

More intriguingly, while nodeless superconducting gaps were observed
in the extremely electron-deopd A$_x$Fe$_{2-y}$Se$_2$ (A = K, Rb,
Cs, ...) with only electron pockets, \cite{YZhang2,DMou,XPWang}
nodal superconducting gap was found in the extremely hole-doped
KFe$_2$As$_2$ with only hole pockets. \cite{JKDong2,KHashimoto3} It
is now under hot debate whether the superconducting gap in
KFe$_2$As$_2$ is $d$-wave or accidental nodal $s$-wave.
\cite{JReid,AFWang1,KOkazaki2} Thermal conductivity measurements
gave compelling evidence for $d$-wave gap, \cite{JReid,AFWang1} but
recent low-temperature angle-resolved photoemission spectroscopy
(ARPES) measurements showed octet-line node structure, suggesting
accidental nodal $s$-wave gap. \cite{KOkazaki2}

To clarify this situation in KFe$_2$As$_2$, it will be helpful to
investigate the superconducting gap structure of its two sister
compounds RbFe$_2$As$_2$ and CsFe$_2$As$_2$, both with $T_c$ = 2.6 K
from the measurements of polycrystalline samples.
\cite{KSasmal,ZBukowski} Unexpectedly, recent muon-spin spectroscopy
measurements of RbFe$_2$As$_2$ polycrystal claimed that the
temperature dependence of the superfluid density $n_s$ is best
described by a two-gap $s$-wave model,
\cite{ZShermadini1,ZShermadini2} which is quite different from
KFe$_2$As$_2$. In this context, more experiments are highly desired,
especially on the single crystals of RbFe$_2$As$_2$ and
CsFe$_2$As$_2$.

In this paper, we present the thermal conductivity measurements of
CsFe$_2$As$_2$ single crystals down to 50 mK. We find clear evidence
for superconducting gap nodes from the significant residual linear
term $\kappa_0/T$ in zero field and the field dependence of
$\kappa_0/T$. Our results suggest common nodal gap structure in
CsFe$_2$As$_2$ and KFe$_2$As$_2$.

\section{Experiment}

The CsFe$_2$As$_2$ single crystals were grown by self-flux method
for the first time. \cite{AFWang2} The sample was cleaved to a
rectangular shape of dimensions 3.5 $\times$ 1.0 mm$^2$
 in the $ab$-plane, with 30 $\mu$m thickness
along the $c$ axis. Contacts were made directly on the sample
surfaces with silver paint, which were used for both resistivity and
thermal conductivity measurements. To avoid degradation, the sample
was exposed in air less than 2 hours. The contacts are metallic with
typical resistance 100 m$\Omega$ at 2 K. In-plane thermal
conductivity was measured in a dilution refrigerator, using a
standard four-wire steady-state method with two RuO$_2$ chip
thermometers, calibrated {\it in situ} against a reference RuO$_2$
thermometer. Magnetic fields were applied along the $c$ axis and
perpendicular to the heat current. To ensure a homogeneous field
distribution in the sample, all fields were applied at temperature
above $T_c$.

\section{Results and Discussion}

\begin{figure}
\includegraphics[clip,width=5.5cm]{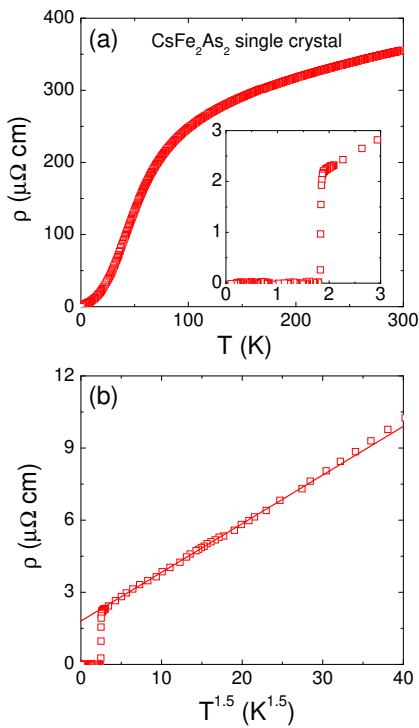}
\caption{(Color online). (a) In-plane resistivity of CsFe$_2$As$_2$
single crystal. The inset shows the resistive superconducting
transition, with $T_c$ = 1.81 K defined by $\rho = 0$. (b)
Low-temperature resistivity plotted as $\rho$ vs $T^{1.5}$. The
solid line is a fit of the data between 2.6 and 9 K to $\rho$ =
$\rho_{0}$ + $AT^{1.5}$, which gives residual resistivity $\rho_{0}$
= 1.80 $\mu$$\Omega$ cm.}
\end{figure}

\begin{figure}
\includegraphics[clip,width=5.5cm]{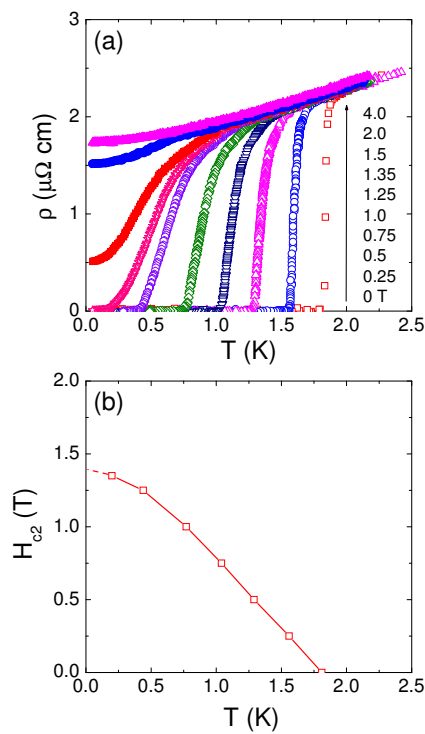}
\caption{(Color online). (a) Low-temperature resistivity of
CsFe$_2$As$_2$ single crystal in magnetic field up to 4 T. (b)
Temperature dependence of the upper critical field $H_{c2}(T)$,
defined by $\rho = 0$. The dashed line is a guide to the eye, which
points to $H_{c2}(0) \approx$ 1.4 T.}
\end{figure}

Figure 1(a) shows the in-plane resistivity $\rho(T)$ of
CsFe$_2$As$_2$ single crystal. The shape of $\rho(T)$ curve mimics
that of KFe$_2$As$_2$ single crystal. \cite{JKDong2,JReid} From the
inset of Fig. 1(a), the $T_c$ defined by $\rho = 0$ is 1.81 K. For
the CsFe$_2$As$_2$ polycrystal in Ref. 34, the $T_c$ defined by the
sharp drop of susceptibility is about 2.2 K. \cite{KSasmal} The
origin of the 0.4 K difference between the $T_c$ of CsFe$_2$As$_2$
single crystal and polycrystal is not clear. One possible reason is
that the polycrystal in Ref. 34 is purer than our single crystal,
since it has been shown that, for KFe$_2$As$_2$, the purer sample
has higher $T_c$. \cite{JReid} In any case, the $T_c$ of (K, Rb,
Cs)Fe$_2$As$_2$ series (3.8 K, 2.6 K, and 1.8-2.2 K, respectively)
seems to decrease with the increase of the ionic radius of alkali
metal.

In Fig. 1(b), the low-temperature resistivity is plotted as $\rho$
vs $T^{1.5}$. It is found that $\rho$ obeys $T^{1.5}$ dependence
nicely above $T_c$, up to about 9 K. The fit of the data between 2.6
and 9 K gives residual resistivity $\rho_0$ = 1.80 $\mu$$\Omega$ cm,
thus the residual resistivity ratio (RRR) = $\rho$(300 K)/$\rho_0
\approx$ 200 is obtained. For the dirty KFe$_2$As$_2$ single crystal
with $\rho_0$ = 3.32 $\mu$$\Omega$ cm and RRR $\approx$ 110, $\rho
\sim T^{1.5}$ has already been noticed. \cite{JKDong2} For the clean
KFe$_2$As$_2$ single crystal with $\rho_0$ = 0.21 $\mu$$\Omega$ cm
and RRR $\approx$ 1180, $\rho \sim T^{1.8}$ was found. \cite{JReid}
Such a non-Fermi-liquid behavior of $\rho(T)$ in KFe$_2$As$_2$ and
CsFe$_2$As$_2$ may result from the antiferromagnetic spin
fluctuations. \cite{SWZhang} In BaFe$_2$(As$_{1-x}$P$_x$)$_2$, the
non-Fermi-liquid linear behavior of $\rho(T)$ near optimal doping,
and the increase of power $n$ in the overdoped regime have been
considered as the signature of a quantum critical point.
\cite{SKasahara}

In order to estimate the upper critical field $H_{c2}$(0) of
CsFe$_2$As$_2$, the resistivity was also measured in magnetic fields
up to $H$ = 4 T, as shown in Fig. 2(a). Fig. 2(b) plots the
temperature dependence of $H_{c2}(T)$, defined by $\rho = 0$. This
definition usually corresponds to the bulk $H_{c2}$. From Fig. 2(b),
we estimate $H_{c2}(0) \approx$ 1.4 T. To choose a slightly
different $H_{c2}$ does not affect our discussion on the field
dependence of $\kappa_0/T$ below.

The ultra-low-temperature heat transport measurement is a bulk
technique to probe the gap structure of superconductors.
\cite{Shakeripour} In Fig. 3(a), we present the temperature
dependence of in-plane thermal conductivity for CsFe$_2$As$_2$
single crystal in zero and applied magnetic fields, plotted as
$\kappa/T$ vs $T$. All the curves are roughly linear, as previously
observed in dirty KFe$_2$As$_2$, \cite{JKDong2}
BaFe$_{1.9}$Ni$_{0.1}$As$_2$, \cite{LDing} and
Ba(Fe$_{1-x}$Ru$_x$)$_2$As$_2$ single crystals. \cite{XQiu}
Therefore we fit all the curves to $\kappa/T$ = $a + bT^{\alpha-1}$
with $\alpha$ fixed to 2. The two terms $aT$ and $bT^{\alpha}$
represent contributions from electrons and phonons, respectively.
Here we only focus on the electronic term.

\begin{figure}
\includegraphics[clip,width=5.5cm]{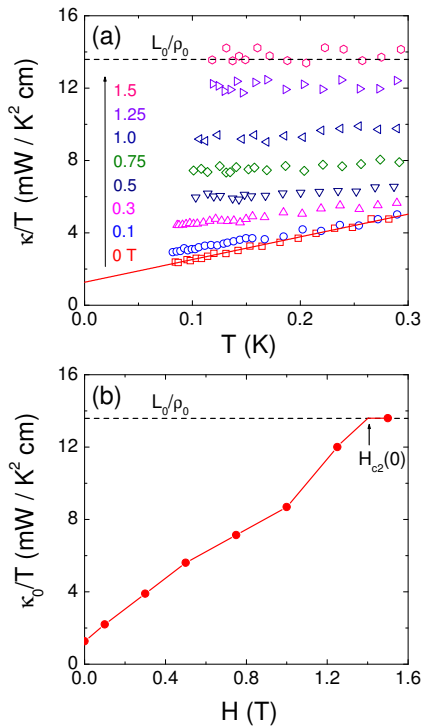}
\caption{(Color online). (a) Low-temperature in-plane thermal
conductivity of CsFe$_2$As$_2$ single crystal in zero and magnetic
fields applied along the $c$ axis. The solid line is a fit of the
zero-field data to $\kappa/T = a + bT$, which gives a residual
linear term $\kappa_0/T$ = 1.27 mW K$^{-2}$ cm$^{-1}$. The dash
lines are the normal-state Wiedemann-Franz law expectation
$L_0$/$\rho_0$, with $L_0$ the Lorenz number 2.45 $\times$ 10$^{-8}$
W $\Omega$ K$^{-2}$ and $\rho_0$ = 1.80 $\mu\Omega$ cm. (b) Field
dependence of $\kappa_0/T$. In $H$ = 1.5 T, slightly above
$H_{c2}$(0) = 1.4 T, the Wiedemann-Franz law $\kappa_0/T$ =
$L_0$/$\rho_0$ is satisfied.}
\end{figure}

\begin{figure}
\includegraphics[clip,width=7cm]{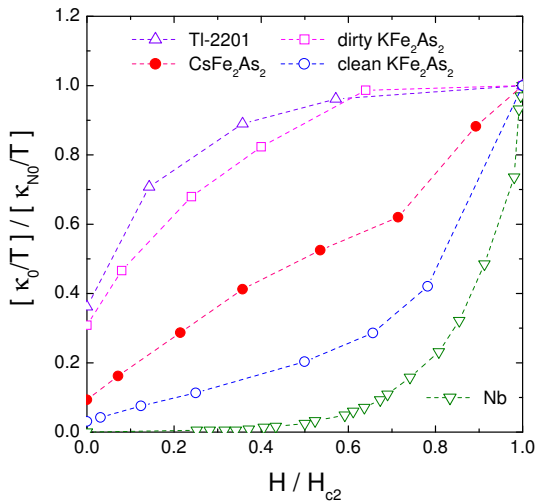}
\caption{(Color online). Normalized residual linear term
$\kappa_0/T$ of CsFe$_2$As$_2$ as a function of $H/H_{c2}$. For
comparison, similar data are shown for the clean $s$-wave
superconductor Nb, \cite{Lowell} an overdoped $d$-wave cuprate
superconductor Tl-2201, \cite{Proust} the dirty and clean
KFe$_2$As$_2$. \cite{JKDong2,JReid}}
\end{figure}

For CsFe$_2$As$_2$ in zero field, the fitting gives $\kappa_0/T$ =
$a$ = 1.27 $\pm$ 0.04 mW K$^{-2}$ cm$^{-1}$. This value is about
1/10 of the normal-state Wiedemann-Franz law expectation
$\kappa_{N0}/T$ = $L_0$/$\rho_0$ = 13.6 mW K$^{-2}$ cm$^{-1}$, with
$L_0$ the Lorenz number 2.45 $\times$ 10$^{-8}$ W $\Omega$ K$^{-2}$
and normal-state $\rho_0$ = 1.80 $\mu\Omega$ cm. For a high-quality
superconductor with no impure phase, such a significant $\kappa_0/T$
in zero field is usually contributed by nodal quasiparticles, thus
considered as a strong evidence for nodes in the superconducting
gap. \cite{Shakeripour} For example, $\kappa_0/T$ = 1.41 mW K$^{-2}$
cm$^{-1}$ for the overdoped cuprate Tl$_2$Ba$_2$CuO$_{6+\delta}$
(Tl-2201), a $d$-wave superconductor with $T_c$ = 15 K.
\cite{Proust} Previously, $\kappa_0/T$ = 2.27 and 3.6 mW K$^{-2}$
cm$^{-1}$ were observed for dirty and clean KFe$_2$As$_2$,
respectively. \cite{JKDong2,JReid}

For the clean KFe$_2$As$_2$, an additional large electronic term
$\kappa/T \sim T^2$ was also observed. \cite{JReid} Reid {\it et
al.} considered this term as the leading-order finite-temperature
correction to $\kappa/T$ for $d$-wave superconductor, which will be
rapidly suppressed by impurity scattering and magnetic field.
\cite{JReid} From Fig. 3(a), such an electronic term is absent in
our CsFe$_2$As$_2$ single crystal, which is not very clean.

The field dependence of $\kappa_0/T$ can provide further support for
the gap nodes. \cite{Shakeripour} For a nodal superconductor,
$\kappa_0/T$ increases rapidly in low field due to the Volovik
effect, \cite{Volovik} as in Tl-2201. \cite{Proust} In contrast, for
a single-gap $s$-wave superconductor, $\kappa_0/T$ displays a very
slow field dependence at low field, as in Nb. \cite{Lowell} In Fig.
3(b), we plot the field dependence of $\kappa_0/T$ for
CsFe$_2$As$_2$. At low field, $\kappa_0/T$ indeed increases rapidly.
Then it shows slight downward curvature before reaching the upper
critical field $H_{c2}$(0). In $H$ = 1.5 T slightly above
$H_{c2}$(0) = 1.4 T, the fitting gives $\kappa_0/T$ = 13.6 $\pm$ 0.3
mW K$^{-2}$ cm$^{-1}$, satisfying the Wiedemann-Franz law perfectly.

For comparison, the normalized $(\kappa_0/T)/(\kappa_{N0}/T)$ of
CsFe$_2$As$_2$ is plotted as a function of $H/H_{c2}$ in Fig. 4,
together with Nb, \cite{Lowell} Tl-2201, \cite{Proust} the dirty and
clean KFe$_2$As$_2$. \cite{JKDong2,JReid} Clearly, the curve of
CsFe$_2$As$_2$ lies between the dirty and clean KFe$_2$As$_2$. The
dirty KFe$_2$As$_2$ shows similar field dependence of $\kappa_0/T$
to that of Tl-2201, \cite{JKDong2} which should also be dirty, with
$\rho_0$ = 5.6 $\mu$$\Omega$ cm and RRR $\approx$ 30. \cite{Proust}
For the clean KFe$_2$As$_2$, Reid {\it et al.} argued that the field
dependence of $\kappa_0/T$ is a compelling evidence for $d$-wave
gap, since the experimental $\kappa_0/T(H)$ curve is close to
calculated curve of a $d$-wave superconductor in the clean limit
($\hbar\Gamma/\Delta_0$ = 0.1). \cite{JReid} The $\rho_0$ and RRR of
our CsFe$_2$As$_2$ lie between the dirty and clean KFe$_2$As$_2$,
indicating that its impurity level lies between the dirty and clean
KFe$_2$As$_2$. This may reasonably explain the position and shape of
its normalized $\kappa_0/T(H)$ curve in Fig. 4. Such a result
suggests that CsFe$_2$As$_2$ has a nodal superconducting gap
structure similar to that of KFe$_2$As$_2$, and shows how the field
dependence of $\kappa_0/T$ evolves with the impurity level.

\section{Summary}

In summary, we have measured the thermal conductivity of
CsFe$_2$As$_2$ single crystal, the sister compound of KFe$_2$As$_2$,
down to 50 mK. Both the significant $\kappa_0/T$ in zero field and
the field dependence of $\kappa_0/T$ provide clear evidence for
nodal superconducting gap in CsFe$_2$As$_2$. Our results suggest
that the extremely hole-doped (K, Rb, Cs)Fe$_2$As$_2$ series of
iron-based superconductors should have a common nodal gap structure.
More experiments on these compounds are needed to get the consensus
on their exact gap symmetry ($d$-wave or accidental nodal $s$-wave).

\begin{center}
{\bf ACKNOWLEDGEMENTS}
\end{center}

This work is supported by the Natural Science Foundation of China,
the Ministry of Science and Technology of China (National Basic
Research Program No: 2009CB929203 and 2012CB821402), and the Program
for Professor of Special Appointment (Eastern Scholar) at Shanghai
Institutions of Higher Learning. \\

$^*$ E-mail: shiyan$\_$li@fudan.edu.cn

\end{document}